\documentclass[%
 reprint,
superscriptaddress,
 amsmath,amssymb,
 aps,
]{revtex4-2}

\usepackage{graphicx}% Include figure files
\usepackage{dcolumn}% Align table columns on decimal point
\usepackage{bm}% bold math
\usepackage{siunitx}
\usepackage[normalem]{ulem}
\usepackage{booktabs}

\newcommand{\alloy}{(Co$_x$Mn$_{1-x}$)$_3$O$_4$}
\newcommand{\CoO}{Co$_3$O$_4$}
\newcommand{\MnO}{Mn$_3$O$_4$}

\begin{document}

\preprint{APS/123-QED}

\title{Free energy of {\alloy} mixed phases from machine-learning-enhanced \textit{ab initio} calculations}

\author{Suzanne K. Wallace}
\affiliation{%
 CEA, LITEN, 17 Rue des Martyrs, 38054 Grenoble, France
}
\affiliation{%
 Universit\'{e} Grenoble Alpes, 621 Avenue Centrale, 38400 Saint-Martin-d'H\`{e}res, France
}
\author{Anton S. Bochkarev}
\affiliation{Interdisciplinary Centre For Advanced Materials Simulation, Universit\"{a}tsstra{\ss}e 150, Ruhr-Universit\"{a}t Bochum, 44801 Bochum, Germany}
\author{Ambroise van Roekeghem}
\affiliation{%
 CEA, LITEN, 17 Rue des Martyrs, 38054 Grenoble, France
}
\affiliation{%
 Universit\'{e} Grenoble Alpes, 621 Avenue Centrale, 38400 Saint-Martin-d'H\`{e}res, France
}

\author{Javier Carrasco}
\affiliation{Centre for Cooperative Research on Alternative Energies (CIC energiGUNE), Basque Research and Technology Alliance (BRTA), Alava Technology Park, Albert Einstein 48, 01510 Vitoria‐Gasteiz, Spain}
\author{Alexander Shapeev}
\affiliation{Skolkovo Institute of Science and Technology, Skolkovo Innovation Center, Nobel St. 3, Moscow 143026, Russia}

\author{Natalio Mingo}%
 \email{Corresponding author: natalio.mingo@cea.fr}
\affiliation{%
 CEA, LITEN, 17 Rue des Martyrs, 38054 Grenoble, France
}
\affiliation{%
 Universit\'{e} Grenoble Alpes, 621 Avenue Centrale, 38400 Saint-Martin-d'H\`{e}res, France
}

\begin{abstract}
{\alloy} is a promising candidate material for solar thermochemical energy storage. A high-temperature model for this system would provide a valuable tool for evaluating its potential. However, predicting phase diagrams of complex systems with \textit{ab initio} calculations is challenging due to the varied sources affecting the free energy, and with the
prohibitive amount of configurations needed in the configurational entropy calculation. 
In this work, we compare three different machine learning (ML) approaches for sampling the configuration space of {\alloy}, including a simpler ML approach, which would be suitable for application in high-throughput studies.
We use experimental data for a feature of the phase diagram to assess the accuracy of model predictions. 
We find that with some methods, data pre-treatment is needed to obtain accurate predictions due to inherently composition-imbalanced training data for a mixed phase. We highlight that the important entropy contributions depend on the physical regimes of the system under investigation and that energy predictions with ML models are more challenging at compositions where there are energetically competing ground state crystal structures.
Similar methods to those outlined here can be used to screen other candidate materials for thermochemical energy storage.

%At different compositions, different entropy contributions dominate the free energy. We find that energy predictions with ML models for this system are more challenging at compositions where there is phase coexistence between different ground state crystal structures.  
%and we provide insights into the most important features to account for in a theoretical model based on the physical regimes of the system of interest.

\end{abstract}

\maketitle

%\tableofcontents

\section{Introduction}

%Accurate prediction of phase diagrams with free energies derived from \textit{ab initio} calculations can be particularly challenging for complex crystal structures. For such systems, it may be necessary to sample extremely large numbers of configurations. 
The mixed phase system {\alloy} is currently under investigation for applications in next-generation concentrated solar energy storage technologies, which are based on reversible redox reactions of metal oxides \cite{Javi_paper, mixed_oxides, Andr2018}. The motivation of the mixed phase is to attempt to minimize the shortcomings of the pure end members, such as cost and toxicity for {\CoO}, and sluggish oxidation rate and poor reversibility for {\MnO} \cite{Javi_13}, by combining them in {\alloy}.
Gaining atomic-level insights into the stability of {\CoO} and {\MnO} mixtures at high temperatures is therefore an important step towards the rational design of these materials. However, the accurate prediction of phase diagrams with free energies derived from \textit{ab initio} calculations can be particularly challenging for complex crystal structures. For such systems, it may be necessary to sample extremely large numbers of configurations.

Based on density functional theory (DFT) calculations of formation energy at $T=$\SI{0}{K}, published in Ref.~\citenum{Javi_paper}, all compositions of {\alloy} should decompose into the pure phases {\CoO} and {\MnO}. However, the mixed phases have been successfully synthesized in a number of works \cite{Javi_paper, mixed_oxides, Andr2018}. This implies that entropic contributions to the free energy of the mixed metal oxide phases must be responsible for the stability at finite temperatures \cite{Widom_rev, entropy_oxides}. It has been suggested that a large number of metastable configurations close in energy to the $T$=\SI{0}{K} ground state may provide the explanation here \cite{Javi_paper}. However, the full configuration space of this mixed phase system would require millions of \textit{ab initio} calculations to study exhaustively, even after discounting symmetrically equivalent structures \cite{Javi_paper}. 
Recent advances in the coupling of artificial intelligence (AI) techniques and  materials design are creating new opportunities to tackle this challenge \cite{ML_MS, ML_MS2, ML_MS3}. 
Specifically, AI-aided approaches, mainly from the subdomain of machine-learning (ML) are helping to significantly reduce the number of required computations, making the materials design process faster and cheaper than conventional high-throughput exploration.

Here in order to investigate possible entropic contributions to the reduction of the free energy of {\alloy}, and to develop a fuller theoretical understanding of the stability of these mixed phases, we use and compare different ML methods to augment the DFT calculations performed in Ref.~\citenum{Javi_paper}.
%The methods developed here could be extended to consider the effect of using other neighbouring species in the periodic table in the mixed phases, such as Cr. This would provide more information for determining mixed phases for optimising the system in terms of an optimal compromise between factors such as: energy storage capacity, price and toxicity.
To evaluate the accuracy of the sampling of the configuration space with these different ML approaches, we use the methods described in Ref.~\citenum{foreground_paper} to calculate the phase coexistence region between the tetragonal hausmannite (H) and spinel (S) phases of {\alloy} from the predicted free energy curves. In each case, the configurational entropy contribution to the free energy is calculated using the different ML approaches and compared to experimental data for this feature of the phase diagram. We discuss the relative strengths of the different methods employed in this work and show that accurate prediction of different parts of the phase diagram of this system depends the most strongly on accurately accounting for different contributions to the free energy.

%\section*{Entropy and thermodynamic stability}\label{thermodynamics}
%Resources: Widom review \cite{Widom_rev}, \cite{Widom_stability}, Box 1 of \cite{HEA_rev}, revisit \cite{HEA_rev2}\\

%\section{Development of the training procedure}\label{training_data}

\section{Mixed metal oxide phases}\label{structure_overview}

%\begin{figure}[h!]
%\includegraphics[width=0.5\textwidth]{figures/crystal_structure_labelled.png}
%\caption{a) cubic spinel (S) structure of {\CoO}, b) and c) tetragonally distorted spinel (hausmannite, H) structure of {\MnO}, shown along different crystallographic planes. Produced with VESTA \cite{vesta}.}\label{crystal_structure}
%\end{figure}

\begin{figure*}
\includegraphics[width=1.0\textwidth]{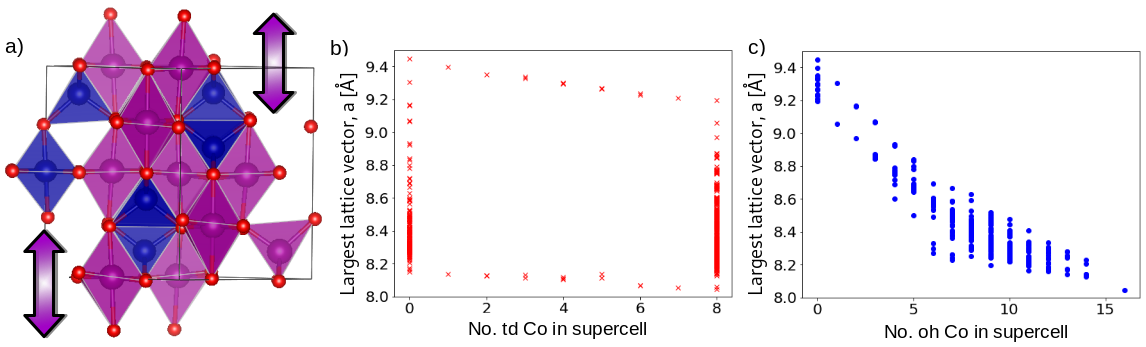}
\caption{a) {\alloy} crystal structure indicating tetragonal distortion introduced by Mn ions on octahedral lattice sites. Produced with VESTA \cite{vesta}. Largest lattice parameter, $a$, in {\alloy} supercells as a function of b) number of Co on tetrahedral (td) lattice sites and c) number of Co on octahedral (oh) lattice sites, data from Ref.~\citenum{Javi_paper} for set A- and B-type structures.}\label{alloy_distortion}
\end{figure*}

The pure {\CoO} phase 
%(shown in Fig.~\ref{crystal_structure}a) 
has a cubic spinel structure (S) with space group 227. Below \SI{1170}{^{\circ}C}, the pure {\MnO} is a tetragonally distorted spinel structure (hausmannite, H) with space group 141 and distortion along one crystallographic axis relative to the S structure. 
%(shown in Fig.~\ref{crystal_structure}b and c). 
Above \SI{1170}{^{\circ}C}, {\MnO} transitions to a S structure \cite{MnO_phases}. The transition metal (TM) sites in these structures are either tetrahedrally (td) or octahedrally (oh) coordinated to the oxygen atoms. 
Mn ions substituted on oh Co sites in the S structure are Jahn-Teller active, inducing a tetragonal distortion of the lattice \cite{CoMnO_expt}. These Jahn-teller active oh sites are largely, but not solely, responsible for a composition-dependent tetragonal distortion of the mixed phases as shown in Fig.~\ref{alloy_distortion}, where small changes in lattice parameters are also correlated to number of td Co. Plots of all DFT lattice parameters as a function of Co composition are shown in the Supplementary information (SI) (Fig. 1).

%\subsection*{Substitution schemes}

 %In the normal spinel structure, AB$_2$O$_4$, the A cations (Co$^{2+}$ or Mn$^{2+}$ in the current system) and the B cations (Co$^{3+}$ or Mn$^{3+}$) preferentially occupy td 8$a$ and oh 16$d$ Wyckoff sites respectively \cite{Javi_paper, Javi_41}. 
 Three distinct Co-Mn substitution schemes were considered in Ref.~\citenum{Javi_paper} and form the basis of the training data used for the ML methods in this study. Taking as a base the pure {\MnO} with the cubic structure of {\CoO}, the data sets referred to as `set A' (`set B') in this study correspond to starting to substitute only td (oh) sites with Co until all td (oh) sites are occupied by Co, and only then beginning to substitute onto the oh (td) sites. Set A and B, as referred to in this work, correspond to `scenario 1' and `scenario 2' respectively in Ref.~\citenum{Javi_paper}. In the data set referred to as `set C' in this work (or `scenario 3' in Ref.~\citenum{Javi_paper}), Co ions are allowed to substitute freely on all TM sites in the crystal. The DFT training data is composed of such structures after atomic positions and cell volumes have been relaxed.
 %In other words, for $N$ Co ions in the supercell, set A configurations in the 56-atom supercell correspond to the formulations Co$_N^{td}$Mn$_{8-N}^{td}$Mn$_{16}^{oh}$O$_{32}$, for $N\leq 8$, and Co$_8^{td}$Co$_{N-8}^{oh}$Mn$_{16-(N-8)}^{oh}$O$_{32}$ for $16>N>8$; set B configurations are in turn formulated as Mn$_{8}^{td}$Co$_N^{oh}$Mn$_{16-N}^{oh}$O$_{32}$, for $n\leq 8$, and Co$_{N-16}^{td}$Mn$_{8-(N-16)}^{td}$Co$_{16}^{oh}$O$_{32}$ for $24>n>16$.\\
 %**Check above + separate equations onto different lines** 
 It is expected that the set C-type structures should be higher in energy than those of either set A or set B types \cite{Javi_paper}, whichever is lower at the concentration of interest. However, without the constraint of preferential occupation, the total possible combination space is dramatically larger, opening up the possibility of entropy-stablisation of the mixed phases \cite{entropy_oxides}. 
 
%\subsection*{Configuration space}
 
 To compare the sampling size required in each of the three substitution scenarios, we take the 56 atom supercell used in the DFT calculations in Ref.~\citenum{Javi_paper}. Each supercell contains 24 TM sites, 8 of which are td-coordinated and 16 are oh-coordinated. In the set C scenario for $N$ Co ions in the supercell, the total number of possible combinations on the 24 TM sites is given by,
 \begin{equation}
     C_N = \frac{24!}{N!(24-N)!} .
 \end{equation}
Whereas in set B, the total number of combinations for N $<$ 16 would be
\begin{equation}
     C_N = \frac{16!}{N!(16-N)!} .
 \end{equation}
Then for N $>$ 16, this would be
\begin{equation}
     C_N =  \frac{8!}{(N-16)!(8-(N-16))!}.
 \end{equation}
 Similarly for set A, the constraint of preferential filling gives a total number of possible combinations for N $<$ 8 as,
 \begin{equation}
     C_N =   \frac{8!}{N!(8-N)!} .
 \end{equation}
Then for N $>$ 8, this would be
\begin{equation}
     C_N =  \frac{16!}{(N-8)!(16-(N-8))!} .
 \end{equation}
Due to the availability of all 24 sites for substitutions in set C-type structures, at intermediate compositions the total number of possible configurations, $C_N$, is enormous. For example, for $x$ = 0.5 in {\alloy}, $C_N$ for set C is 2.70$\times10^6$. In comparison, for the same $x$ for sets A and B, $C_N$ is just 1820. For this reason, the data set in Ref.~\citenum{Javi_paper} sampled sets A and B thoroughly, but was unable to substantially sample set C, even after eliminating symmetrically equivalent structures. Therefore, it was not possible in this study to consider configurational entropy contributions from this large portion of the total configuration space. For this reason, we have used the DFT data set of Ref.~\citenum{Javi_paper} as a starting point for training different ML models to predict the energies of the set C-type structures.

\section{Machine learning methods}\label{ML_methods}

Recent years have seen many successes in the use of ML methods for various applications in materials science \cite{ML_rev_gen, ML_rev_gen2, ML_QM_rev}, where data sets are used to train algorithms to predict properties of interest. 
Here, to approximate the formation energy in complex {\alloy}  mixed phases, we apply supervised machine learning~\cite{python_ML}.
%There are different types of ML, such as: supervised, unsupervised and reinforcement learning \cite{python_ML}. Here we have used supervised learning for a regression task, where data is used to determine the form of a function for mapping from input variables to a continuous output variable. This functional form can then be used to predict output variables for `unseen' data based on its input variables. In the case of this study, and numerous others, the input variables are the atomic configuration of the system and the output variable is the total energy of the given configuration. 
For this, %particular application, 
various learning algorithms %have been successfully applied 
and various representations for the atomic configurations (or `descriptors') 
%have been developed 
exist %
which take into account rotational and translational invariance \cite{Behler_NN, MTP_orig, GAP, SOAP, Legrain2017}. Comparisons of some of these methods can be found, %in works such as Ref.~\citenum{surrogate} and \citenum{MTP_GAP_transition}, %where it has been demonstrated that for the different methods there can be 
highlighting their
differing predictive capabilities and computational cost \cite{surrogate,MTP_GAP_transition}.

Common to all of these methods is the need to optimise the model. 
%, and for this there are numerous ways to tune the training procedure. 
During training it is important to consider overfitting and underfitting, %also referred to as the 
or `bias-variance' trade off~\cite{ML_rev}. 
%A high bias (or underfitting) model is too simple or inflexible to describe well the relationship between the input representation of the training data and its corresponding output variable. Whereas a high variance (or overfitting) model is able to fit very well to the training data, but performs poorly when making predictions for unseen data \cite{ML_rev}. 
To avoid overfitting or underfitting, it is common to trial various `hyperparameters' associated with the model complexity for the particular ML method and to then compare the root mean square error (RMSE) between reference outputs and model predictions at the end of the training process. This is done with both the training and validation datasets, to optimize the hyperparameters, and a `hold-out' set of data, which was not seen by the algorithm during training, to assess model accuracy. Note that some differing terminologies are used in the literature where the term ``validation set" is sometimes used for the ``hold-out set", or is sometimes replaced by ``test set". In the case of underfitting, the RMSE on the training data is large. While for overfitting the RMSE on the training data is typically small, but coincides with either a plateau or increase in the RMSE for the validation set with increasing model complexity. 

In this work, we investigate the use of three different supervised ML methods. The models are trained with a data set of hundreds of final, relaxed structures of {\alloy} and their corresponding total energies. 
%The models are then used to predict the total energies of set C-type structures. 
In the next sections we provide a brief outline of each method and a description of their corresponding training hyperparameters.

\subsection{Artificial neural networks (ANNs)}

Our ANN methodology is that of Ref.~\citenum{Anton_NN}, which makes use of symmetry functions for representing chemical environments, similar to those developed by Behler and Parrinello \cite{Behler_NN, Behler_Parrinello, Behler_tutorial}. During the construction of a descriptor for the chemical environment, it is important to ensure that it is invariant with respect to rotations and translations of the system and also to permutations in the ordering of the atomic inputs. The design of the descriptor can also allow for more automatic optimisation of the model, and hence fewer hyperparameters that need to be tuned manually \cite{Anton_NN}. 

The representation of input data to the ANN in this study is an $n$-dimensional vector for each atom $i$ in the system,
\begin{equation}
    D^l_i = \sum c^l(z_i)c^l(z_j)\mathrm{e}^{-\sigma^l(r_{ij} - \eta^l)^2} f_{\mathrm{cut}}(r_{ij}, R_{\mathrm{cut}}) ,
    \label{ANN_descriptor}
\end{equation}
which contains information about the atoms surrounding atom $i$ out to a certain cutoff distance, $R_{\mathrm{cut}}$, by using a Gaussian expansion of interatomic distances $r_{ij}$ to probe the surrounding space where $l$ labels each Gaussian in the expansion. Parameters $\eta$ and $\sigma$ are related to the probing of this space \cite{Anton_NN}. In this particular implementation, these two parameters are automatically optimised during training so that regions of space which are more important for distinguishing different atomic configurations are selected. $f_{\mathrm{cut}}$ is the cutoff function which ensures that the contributions from neighbouring atoms smoothly goes to zero as $r_{ij}$ approaches $R_{\mathrm{cut}}$. $c^l(z_i)$ and $c^l(z_j)$ are the parts of the vector descriptor that are related to the chemical identity of atom $i$ and its neighbour $j$, defined by their atomic numbers $z_i$ and $z_j$. These vectors are initialised randomly and are also optimised automatically during training.

The output of this ANN is the energy of atom $i$. The total energy of a system is then the summation of all of the atomic energies,
\begin{equation}
 E_{\mathrm{tot}}^{\mathrm{ANN}} = \sum_i E_i  .
\label{ML_Etot}
\end{equation}
The ANN is trained by minimising the loss function. The original implementation in Ref.~\citenum{Anton_NN} and also in Ref.~\citenum{Anton_NN_nanowire} trains simultaneously with forces and energy, whereas in this study we train only with energies due to the lack of force information in the initial data set. Without including the forces, the loss function in Ref.~\citenum{Anton_NN} is reduced to
\begin{equation}
\big\| E_{\mathrm{tot}}^{\mathrm{ANN}} - E_{\mathrm{tot}}^{\mathrm{DFT}} \big\|^2 ,
    \label{ANN_loss}
\end{equation}
where $E_{\mathrm{DFT}}$ are the reference total energies of the training data and $E_{\mathrm{ANN}}$ are the predictions by the ANN model for the same atomic configurations. Essentially, the ANN model is fit to minimise the RMSE between DFT energies and predicted energies for the training data.

Despite the automatic optimisation of some of the parameters in Eq.~\ref{ANN_descriptor}, there are still various hyperparameters that need to be tuned for the ANN, and carefully chosen to avoid overfitting or underfitting. Firstly, there is the cutoff distance, $R_{\mathrm{cut}}$, used when creating the descriptor for each atom in the system. Then for model complexity, the architecture of the neural network can be tuned by varying the layer size and depth. A larger network allows for more model flexibility, but too large can result in overfitting to the training data. Lastly, the number of epochs can also be tuned. This is the number of training cycles the model is optimised over. Again, here if the model is trained over too many epochs this can result in overfitting.

\subsection{Moment tensor potentials (MTPs)}\label{MTP_overview}

MTPs are a class of ML potentials first proposed in Ref.~\citenum{MTP_orig} and later extended for multi-component systems \cite{MTP_many}. The latest release of the software is described in Ref.~\citenum{MLIP_v2}. Similarly to the ANN methodology outlined above, MTPs represent the energy of an atomic configuration as a sum of contributions of local atomic environments of each atom $i$,
%. The total energy of the configuration is obtained from the potential energy of interatomic interaction summed over each atom,
\begin{equation}
    E_{\mathrm{tot}}^{\mathrm{MTP}} = \sum_{i=1}^n E_i.
\end{equation}
Each contribution $E_i$ is linearly expanded via a set of basis functions, $B_{\alpha}$:
\begin{equation}\label{MTP_basis}
    E_i = \sum_{\alpha} \xi_{\alpha} B^{\alpha}_i %(|r_{ij}|, z_i, z_j) ,
\end{equation}
where the set of parameters $\boldsymbol{\xi} = \{ \xi_{\alpha} \}$ are obtained during training.
Unlike ANNs, the atomic environments are instead represented by the moments of inertia of the neighbouring atoms.
The moment tensor descriptors of the $i$th atom consist of radial and angular parts and are given by:
\begin{equation}\label{MTP_descriptor}
    M^{\mu, \nu}_i = \sum_j f^{\mu}(r_{ij}, z_i, z_j) \underbrace{\mathbf{r}_{ij} \otimes \ldots \otimes \mathbf{r}_{ij}}_\text{$\nu$ times}.
\end{equation}
Here $\mathbf{r}_{ij}$ is the position of the $j$th atom relative to the $i$th atom and $r_{ij}$ is, as before, its length.
The radial part is further expanded as
\begin{equation}\label{MTP_radial}
     f^{\mu}(r_{ij}, z_i, z_j) = \sum_{\beta = 1}^{N_Q} c_{\mu, z_i, z_j}^{(\beta )} Q^{(\beta )}(r_{ij}) .
\end{equation}
and defines different shells around atom $i$ and contains a set of radial parameters $\mathbf{c} = \{c_{\mu, z_i, z_j}^{(\beta )} \}$ which are also obtained during training, and radial basis functions, $Q^{(\beta )}(r_{ij})$ based on polynomials. A cutoff radius, $R_{\mathrm{cut}}$, is also used here to ensure smooth behaviour at the edges of the atomic environments.
The angular part, $\underbrace{\mathbf{r}_{ij} \otimes \ldots \otimes \mathbf{r}_{ij}}_\text{$\nu$ times}$, contains the angular information about the atomic environment and is a rank $\nu$ tensor. 
%When $\nu = 0$, the angular component is a scalar. For $\nu=1$ this component is a vector pointing from atom $i$ to atom $j$. For $\nu = 2$, this component has a matrix form. 
When the radial and angular components are combined to form $M^{\mu, \nu}$, for $\mu = 0$, the increasing ranks of the tensor can be interpreted mechanically as: the number of atoms within $R_{\mathrm{cut}}$ of atom $i$ (or as the `mass' of these atoms) for $M^{0,0}$, the centre of mass scaled by the mass for $M^{0,1}$ and the tensor of the second moments of inertia for $M^{0,2}$, etc. For $\mu > 0$, this can then be interpreted as weighted moments of inertia \cite{MTP_orig}.

The basis functions, $B_{\alpha}$, in Eq.~\ref{MTP_basis} are constructed by defining the `level of moments' via,
\begin{equation} \label{MTP_levels}
    \mathrm{lev} M^{\mu, \nu} = 2 + 4\mu + \nu,
\end{equation}
where the coefficients in Eq.~\ref{MTP_levels} were found to be optimal in Ref.~\citenum{MTP_HT}. The tensor contractions of a number of moments are defined by adding together such levels. All such contractions of one or more moments form the basis functions, $B_{\alpha}$. As for the symmetry functions representation with the ANN, these basis functions are invariant to atomic permutations, rotations and reflections.
The functional form of the MTP is defined firstly by choosing a maximum level for the basis set, $\mathrm{lev}_{\mathrm{max}}$, and then including all basis functions whose level is less than or equal to that maximum. And secondly, by the size of the radial basis, $N_Q$, in Eq.~\ref{MTP_radial}.

The parameters $\boldsymbol{\xi}$ and $\mathbf{c}$ make up the total set of parameters that are found during training, $\boldsymbol{\theta} = \{ \boldsymbol{\xi}, \mathbf{c} \}$. 
The total number of basis functions (and hence the number of the corresponding parameters $\boldsymbol{\xi}$) grows exponentially with $\mathrm{lev}_{\mathrm{max}}$, but the number of radial functions increases linearly with $\mathrm{lev}_{\mathrm{max}}$ and $N_Q$. $\mathrm{lev}_{\mathrm{max}}$ and $N_Q$ are therefore the hyperparameters which define the total set of parameters to be found during training and hence define the model complexity. The computational expense increases with the total number of free parameters; and the optimal number of such parameters depends on the total training set size, with the possibility of overfitting with too many parameters for a small data set.
During the training process, fitting with MTP is performed using output variables  for each configuration in the training set for quantities from \textit{ab initio} calculations: total energy, forces and the stress tensor. Weights can be set to express the importance of each of these quantities during the optimisation. In the case of this work, as the initial data set did not contain sufficient force information, weights for forces were set to zero.

\subsection{Extended Mean Field (EMF) model for a two-species, two-site-type compound}

An approach commonly known as the Miedema model is often used to describe the mixing enthalpy of alloys \cite{Miedema_orig, Widom_rev}. The Miedema model was originally developed in the context of liquid alloys, and it was expressed in terms of solution enthalpies and interfacial area between species. Nevertheless, its underlying basic concept is shared with a mean field model in which the average energy of atomic configurations with average species occupations $\phi_i$ is given by atom-pair contributions, $\phi_i \epsilon_{i,j} \phi_j$, where $i$ and $j$ denote the species type \cite{provatas2010phase}. In binary mixtures of species $a$ and $b$ the two concentrations are related by $\phi_b = 1-\phi_a$. The total energy is
\begin{equation}
\langle E\rangle =\phi_a^2 \epsilon_{aa} + 2\phi_a \phi_b \epsilon_{ab} + \phi_b \epsilon_{bb}.
\end{equation}

The case of {\alloy} is more complex because there are two distinct site types: oh and td, and two species: Co and Mn. We thus extend the model to account for all the possible interactions between pairs of species and site types. There are four types of atoms: Co on a td site, Co on an oh site, Mn on a td site, and Mn on an oh site. This results in a 4x4 matrix of pair interactions $\epsilon_{ij}$. However, the matrix is symmetric because $\epsilon_{ij}=\epsilon_{ji}$, so only the 10 elements contained in the diagonal and the upper triangle are independent parameters.
We can write the energy as the quadratic form:
\begin{equation}
    \langle E \rangle = \underline{x} \mathbf{\epsilon} \overline{x} \\
    = x_i \epsilon_{ij} x_j,
\end{equation}
where we assume summation over repeated indices.
Here $\underline{x}=\overline{x}^{T}=\{x_{\mathrm{td}},1-x_{\mathrm{td}},x_{\mathrm{oh}},2-x_{\mathrm{oh}}\}$ defines the average occupations of the system, where $0\leq x_{\mathrm{td}} \leq 1$ ($0\leq x_{\mathrm{oh}}\leq 2$) denote the average Co concentration in the td (oh) sites, and $1 - x_{\mathrm{td}}$ and $2 - x_{\mathrm{oh}}$ denote the corresponding Mn concentrations. The occupation range of the oh sites is twice as large as for the td sites, since there are twice as many of the former than of the latter. The $\epsilon_{ij}$ are adjustable parameters to be fitted to the DFT energies. To do this, we reshape the 10 independent elements of $\mathbf{\epsilon}$ as a single column array, $\overline{M}$, and write $N$ equations for each of the DFT configurations that we want to include in the fit, as
\begin{equation}
    E^{(n)} = x^{(n)}_i x^{(n)}_j M_{u} = A_{n,u} M_u,
    \label{eq:miedema_matrix_form}
\end{equation}
where $u \equiv (i,j)$ labels the ordered list of pairs $(i,j)$, with $1\leq i \leq 4$, $1\leq j \leq i$, containing a total of 10 elements. $E^{(n)}$ is the DFT energy corresponding to configuration $n$, whose concentrations array is $\overline{x}^{(n)}=\{x^{(n)}_{\mathrm{td}},1-x^{(n)}_{\mathrm{td}},x^{(n)}_{\mathrm{oh}},2-x^{(n)}_{\mathrm{oh}}\}$. Directly substituting this into Eq.~\ref{eq:miedema_matrix_form}, the explicit values of $\overline{A_n}$ in terms of the elements of $\overline{x}^{(n)}$ are 
\begin{equation}
\begin{aligned}
\overline{A} = \{x_{\mathrm{td}}^2, (1 - x_{\mathrm{td}}) x_{\mathrm{td}}, x_{\mathrm{oh}} x_{\mathrm{td}},
(2 - x_{\mathrm{oh}}) x_{\mathrm{td}},\\ (1 - 
  x_{\mathrm{td}})^2,  x_{\mathrm{oh}} (1 - x_{\mathrm{td}}), (2 - x_{\mathrm{oh}}) (1 - x_{\mathrm{td}}),\\
  x_{\mathrm{oh}}^2, (2 - x_{\mathrm{oh}}) x_{\mathrm{oh}}, (2 - x_{\mathrm{oh}})^2 \},
\end{aligned}
  \label{eq:A_elements}
\end{equation}
where the superscript $(n)$ is ommitted from the $x$ for simplicity.

Eq.~\ref{eq:miedema_matrix_form} is a system of $N$ equations with 10 unknowns. It can be written in compact form as 
\begin{equation}
    \mathbf{A}\overline{M}=\overline{E}.
\end{equation}
The solution to this (usually overdetermined) system can be obtained as
\begin{equation}
    \overline{M} = \mathbf{A}^{-1} \overline{E},
    \label{eq:solution}
\end{equation}
where $\mathbf{A}^{-1}$ is the pseudoinverse of $\mathbf{A}$. We implement this solution via a standard least-squares algorithm.
Rather than fit to the total DFT energies, the model is fit with the formation energies calculated via 
\begin{equation}\label{form_E}
    E^{(n)}_{\mathrm{form}}\equiv E^{(n)}_{tot}(x) - \left ((1-\frac{x}{3})E^{x=0}_{tot} - \frac{x}{3}E^{x=3}_{tot}\right), 
\end{equation}
where $E^{x=0}_{tot}$ and $E^{x=3}_{tot}$ correspond to the total energy of the pure phases. 
Formation energies are zero at the composition extremes, when $x =0$ or $x = 3$. To ensure that this is the case, from each equation $(n)$ in Eq.~\ref{eq:miedema_matrix_form} we subtract the energy corresponding to the linearly weighed average between $E(x=0)$ and $E(x=3)$ at $x(n)$. Using Eq.~\ref{eq:A_elements}, this yields 
\begin{equation}
    \overline{E}_{\mathrm{form}}^{(n)} = \left (\overline{A}_{n}  - \frac{x^{(n)}_{\mathrm{td}} + x^{(n)}_{\mathrm{oh}}}{3}\overline{B} - \overline{C} \right )\overline{M},
    \label{eq:miedema_matrix_form2}
\end{equation}
where $\overline{B} =  \{1, 0, 2, 0, -1, 0, -2, 4, 0, -4\}$ and $\overline{C}=\{0, 0, 0, 0, 1, 0, 2, 0, 0, 4\}$.
This equation is solved similarly to Eq.~\ref{eq:solution}.

The only means to improve the prediction performance of the EMF model is to increase the training set size, compared to the various different hyperparameters associated with training the ANN or MTP. Unlike the ANN and MTP models, the implementation of the EMF model in this work does not depend on the exact atomic positions or cell volume of the relaxed structures. While the classic Miedema model involves additional terms, such as elastic strain and lattice structure information \cite{Widom_rev, Miedema_eg1, Miedema_eg2}, the EMF model in this work depends only on the oh or td site occupancy of the Co or Mn species.
%Therefore, it will be unable to distinguish energies of structures with the same number of Co and the same oh:td ratio.

\section{Results and discussion}

\subsection{ML prediction of formation energies}
\subsubsection{Development of the training procedure}
\subsubsection*{Training, validation and hold-out sets}\label{training_data}

The initial DFT training set from Ref.~\citenum{Javi_paper} contains 540 relaxed structures of A- and B-type and 16 C-type (see Section~\ref{structure_overview}).
Another 496 C-type structures were calculated with the same settings as in Ref.~\citenum{Javi_paper}: spin-polarised DFT calculations with the Vienna ab initio simulation package (VASP) \cite{VASP1, VASP2} and the strongly constrained and appropriately normed (SCAN) meta-generalized gradient approximation (meta-GGA) functional \cite{SCAN}.
115 of the C-type structures contained 12 oh Co and 2 td Co. %This was to allow testing of the predictive power of the ML models for data sets with fewer distinguishing features. 
The remaining 381 C-type structures calculated are randomly selected with various compositions. The total DFT data set contains 1052 structures. When predicting energies for the full configuration space of set C, the entire DFT data set described above is used to train and validate the ML model. However, the design of the training procedure (outlined in the following subsections below) involved splitting this full training set into various training, validation and hold-out sets.

\subsubsection*{Imbalanced data sets}

\begin{figure*}[ht]
\includegraphics[width=0.8\textwidth]{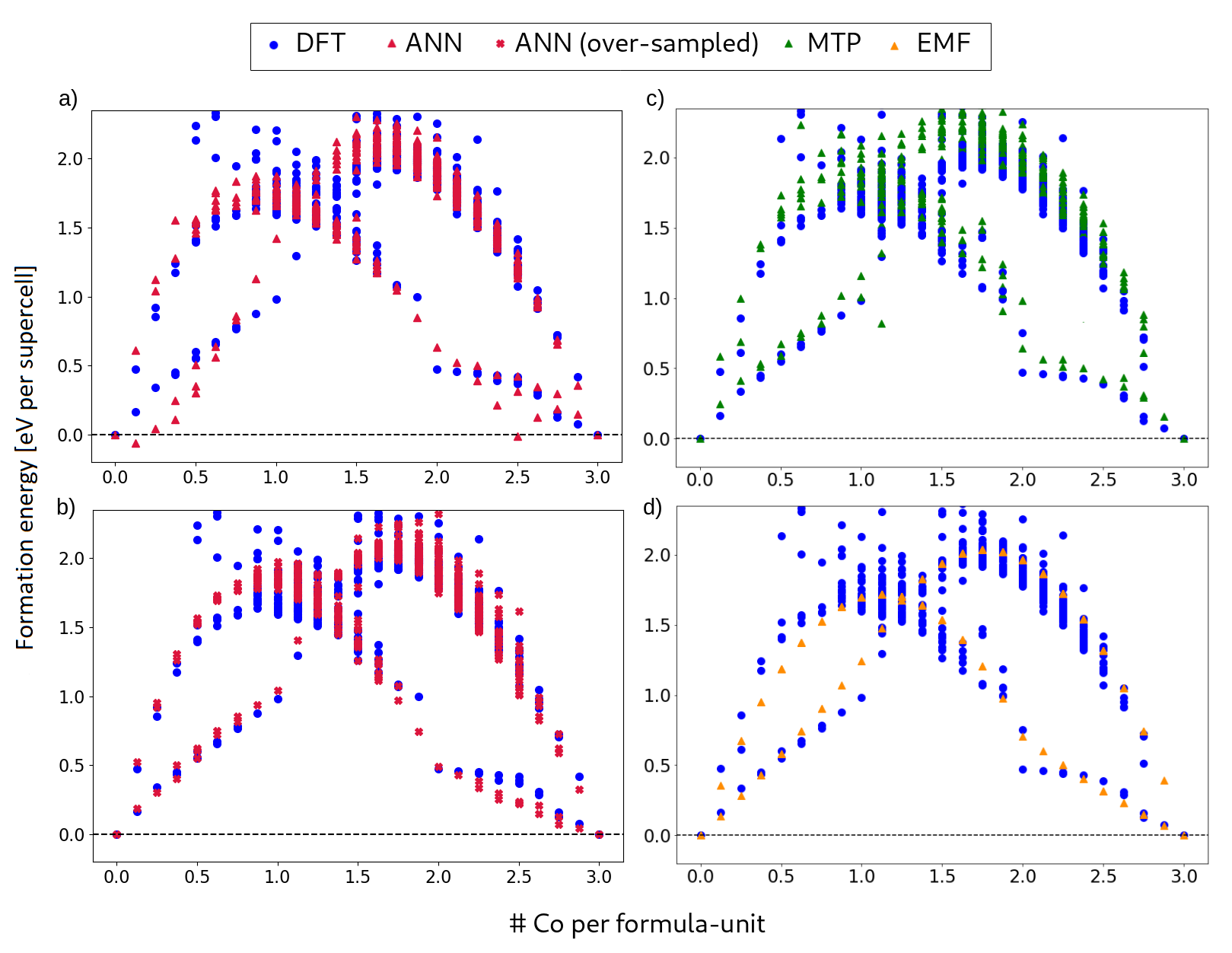}
\caption{Predicted formation energies and reference DFT data for set A and B type configurations of {\alloy}. All models are trained with 360 structures of the total data set shown (540 structures). Structures not included in the training set are used as the hold-out set to calculate the RMSE in units of meV per atom. a) Predictions from the ANN with a RMSE of 3.1. b) Predictions from the ANN with an `over-sampled' training set and a RMSE of 3.6. c) Predictions from the MTP method with a RMSE of 4.6. d) Predictions from the EMF model with a RMSE of 2.7.}\label{imbalanced_hull}
\end{figure*}

The training data across the full composition range contains many more possible structures at intermediate compositions than at the composition extremes. This can pose a challenge when wanting to accurately predict energies of structures that are less well-represented in the training data. As a test, the DFT data for A- and B-type structures from Ref.~\citenum{Javi_paper} was randomly split into a training set with 360 structures, and a test set with the remaining ones.
%, where the latter is not included in the training data but is used to compute the RMSE between energy predictions and reference energies from DFT. 
%360 structures were randomly selected to be the training set for all three models and the remaining structures were used as the test set.
%This test was performed using unrelaxed structures with the corresponding relaxed total energies, as the ability to predict the relaxed total energy for an input unrelaxed structure would provide a much more useful and powerful predictive tool for this study.
The initial training data for set A- and B-type structures contained only the final, relaxed structures. Therefore, there was insufficient force information from the relaxation trajectory to fit interatomic potentials for the systems, as is often done in other ML studies to predict the full potential energy surface (PES) \cite{GAP_PES, Behler_perspective, Behler_NN, Anton_NN}. Only the basins of the PES for each mixed phase were of interest for this study. Therefore, instead ML models were trained to map initial, unrelaxed atomic configurations onto final, relaxed energies based on DFT data of similar structures. 

The different hyperparameters associated with each of the ML methods (see Methods section) were trialled to determine those which resulted in predictions for the validation set with the smallest RMSE. 
The best-performance models were then used to predict total energies, and from them formation energies, $E^{(n)}_{\mathrm{form}}$ via Eq.~\ref{form_E}.
Plots of formation energy as a function of composition are shown in Figs.~\ref{imbalanced_hull}a-c.
The ANN method yields a smaller RMSE than MTP (3.1 meV per atom vs. 4.6 meV per atom). However,  it performs worse at the composition extremes than MTP, where the formation energy even drops below the dashed line marking zero energy at certain compositions.
%We investigated different sampling methods for the imbalanced data set with the ANN. The data set was not sufficiently large to simply randomly under-sample the majority data at intermediate compositions. Instead, 

In order to better balance the sampling for the ANN training,
we trialled generating artificial data for the minority data based on k-nearest neighbours using the SMOTE algorithm \cite{SMOTE}. However, the structures that this method generated were often unphysical. This is likely to be due to the small sample size and high-dimensionality of the data with each atom type, its coordinates as well as the types and coordinates of its neighbours within the cutoff radius all being used to define the representation of the structure. 
We found that the predictive performance of the ANN at composition edges improves by weighting higher (or `over-sampling') the minority data during training (Fig.~\ref{imbalanced_hull}b). This however slightly increases the RMSE from 3.1 to 3.6 meV per atom. 

Despite the low RMSE's obtained with the ANN, based on the better qualitative agreement of the MTP predictions for this imbalanced data set without the need for any data pre-treatment, we do not proceed further with the ANN method for this study.
The optimal pre-treatment of training data is an open research question \cite{imbalanced_classification_ML} and beyond the scope of this work. However, the improvements observed from over-sampling imply that further investigations into data pre-treatment procedures could result in a very good predictive performance with the ANN.
The optimal choice of ML method and data preparation procedure may also be dependent upon the particular system under investigation and the available data. For example, the prediction accuracy may differ considerably for a large, balanced data set where it is desirable to reduce the impact of outliers in the data set on the model. %The sole purpose in the training procedure of our ANN is to minimise the RMSE between predictions and the reference labels for this data, with no way of weighting data based on its rarity. In the case of the energies across the full composition range in this system, this resulted in very good predictions at compositions with the most data, but relatively poor performance at composition extremes with few data points. However, typically better RMSE values were obtained. It is possible that an ANN may be the better method for a large, balanced data set where it is desirable to reduce the impact of outliers in the data set on the model.
 
The third method trialled is the EMF model shown in Fig.~\ref{imbalanced_hull}d. This method gave the best RMSE of all three methods (2.7 meV per atom) and did not suffer from the visibly poor predictions at the composition edges as with the ANN. It is likely that applying the constraint of zero formation energy at the composition extremes in the EMF model is responsible for this improvement. However, the EMF predictions do not reproduce well the spread in the DFT data at each composition.

\subsubsection*{Assessment of prediction capabilities}\label{12oh2td}

\begin{figure}[h!]
\includegraphics[width=0.475\textwidth]{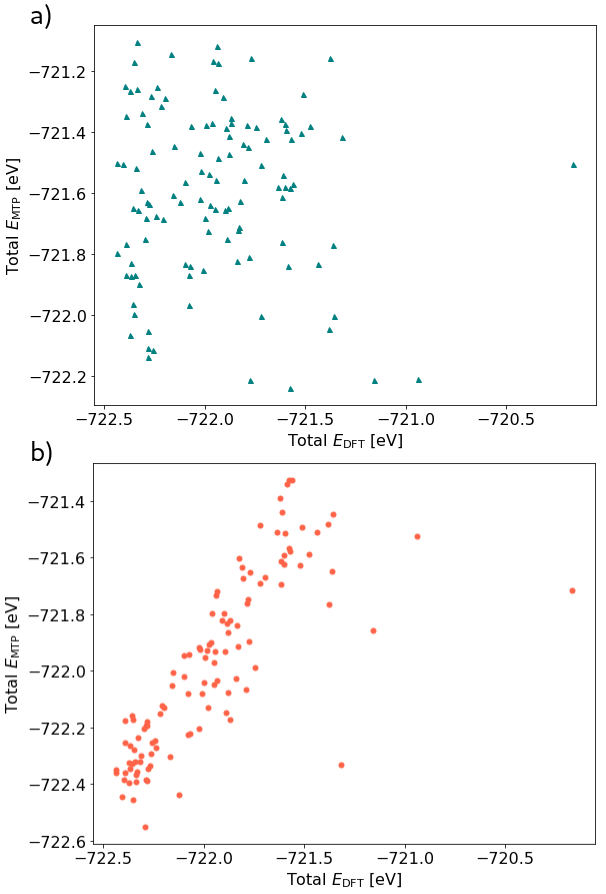}
\caption{Correlation between reference DFT energies and MTP predicted energies of the set C 12 oh 2 td hold-out set with a) completely unrelaxed structures as inputs and b) structures with unrelaxed ionic coordinates scaled by relaxed lattice parameters.}\label{Pearson_coeff}
\end{figure}

Fig.~\ref{alloy_distortion} shows a substantial variation in the lattice parameters of the {\alloy} supercells as a function of the number of Co. It has been shown that cluster expansion predicted energies are typically less accurate when the systems undergo substantial atomic relaxation \cite{CE_relaxation}.
In the data set for {\alloy}, the total energy of each supercell is most strongly dominated by the total number of the substituting species, i.e. number of Co, in the supercell. However, finer energy differences between structures depend on the fraction of Co on td or oh sites and, even more subtly, the variations in atomic arrangements between structures with the same number of Co on the same type of crystallographic sites.
As our EMF model takes only the number of oh and td Co as its representation of the structure, it is unable to distinguish between different structures having the same number of oh and td Co.

The ANN and MTP methods are able to distinguish the energies of structures with the same number of oh and td Co only if the unrelaxed ionic coordinates in the input structures are scaled by the relaxed lattice parameters.
%This was tested by training the model with all of the DFT data, except for a test set of 115 set C-type structures, all with 12 Co on oh sites and 2 on td sites. 
%To assess predictive performance, the RMSE for the test set and the Pearson correlation coefficient between test set reference DFT energies and the corresponding predictions are calculated. 
Fig.~\ref{Pearson_coeff} shows the predicted MTP versus DFT energies when using a completely unrelaxed structure as the input (a), and when the unrelaxed ionic coordinates are scaled by the relaxed lattice parameters (b). The Pearson correlation coefficient increases from -0.0398 to 0.778 and the RMSE decreases from 6.14 to 0.969 meV per atom when using the relaxed lattice parameters. Therefore it is important to obtain estimates of relaxed lattice parameters for set C structures before predicting their total energy with the MTP model, as described in the next subsection.

Ref.~\cite{CE_relaxation} highlighted the crucial influence of atomic relaxation in the accuracy of the predicted formation energies of mixed phases using cluster expansion. It appears that a similar phenomenon is at play in our particular use of ML i.e. training only on fully relaxed configurations, whereby strong relaxations lead to decreased prediction capability. Nonetheless, this does not prevent the accurate prediction of the solubility gap (section~\ref{free_energy_pred}), or to calculate specific site Co occupation probabilities that are in principle verifiable by experiment~\cite{supplemental-information}. The fact that the training set is one order of magnitude smaller than one would need to train a full-fledged interatomic potential makes this approach useful and attractive when the availability of computed ab initio data is limited.

\subsubsection*{Prediction of relaxed lattice parameters}\label{lattvec_pred}

\begin{figure*}
\includegraphics[width=1.0\textwidth]{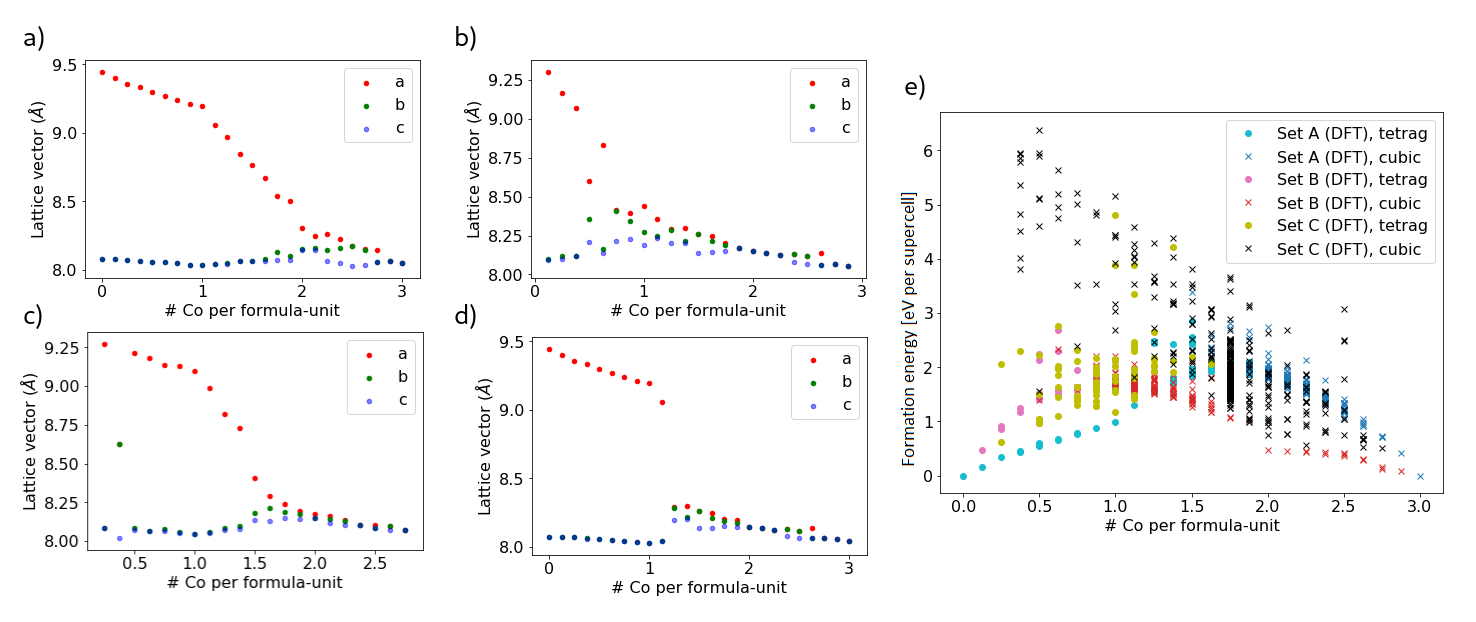}
\caption{Lattice parameters (denoted $a$-$c$ from largest to smallest) as a function of the number of Co atoms in the supercell from the training DFT data for the lowest-energy structures that are a) A-type, b) B-type, c) C-type, d) all three sets combined. Structures are identified as cubic when all lattice parameters are equal, within a tolerance of 0.05{\AA}. e) Formation energy as a function of composition for the same DFT data with structures identified as cubic (tetragonal) denoted by crosses (filled circles).}\label{phase_transition}
\end{figure*}

Studies on alloys often use Vegard's \cite{surrogate, vegard} empirical law of a linear relationship at constant temperature between the lattice parameters and the alloy concentration, resulting from the different sizes of the substituting atoms. However, for {\alloy}, factors other than concentration also affect the lattice parameters for a given Co:Mn ratio. The occupation of the oh sites strongly influences the lattice parameters (section \ref{structure_overview}). Also, increasing Mn content does not smoothly distort the cubic structure into the tetragonal cell shape of {\MnO}. DFT training data in Fig.~\ref{phase_transition}d, and experimental measurements for this system \cite{CoMnO_expt} show a composition-dependent phase transition of the ground state structure from the tetragonal spinel of {\MnO} to the cubic spinel of pure {\CoO}. The composition at which the ground state phase transition occurs differs between the A-, B- and C-type structures (Fig.~\ref{phase_transition}a-c).

The lack of force information in the initial training data set prevents us from relaxing cell shapes with an MTP potential. Instead, we use all of the DFT training data (section \ref{training_data}) to estimate lattice parameters based on the number of oh and td Co in the supercell. The training data is classified as cubic when all three lattice parameters are the same within a tolerance of 0.05{$\AA$}, and tetragonal otherwise. This gave 824 cubic  and 228 tetragonal structures. Their formation energies as a function of composition are displayed in Fig.~\ref{phase_transition}e. 
Best-fit quadratic surfaces are then fit to cubic and tetragonal data separately for each lattice parameter as a function of the number of Co on td and oh sites in the supercells. For cases with several structures with the same number of oh and td Co, the minimum energy structure is used in the fit. These surfaces are then used to predict the lattice parameters for cubic and tetragonal structures, both when sampling the full configuration space of set C, and also to replace the true relaxed lattice parameters of the training set. All best-fit surfaces are included in the SI (Sections 2 and 3). 

For both tetragonal and cubic structures, the largest weighted mean error in the fits was for the largest lattice parameter, $a$, which is the one varying the most as a function of composition. With this method any structure with the same number of oh and td Co will be assigned the same volume. However, the distribution of cell volumes in the DFT data for C-type structures shows substantial variations, even between structures with the same number of oh and td Co (see Section 4 of the SI). Cell volumes of the 12 oh 2 td Co set have a range of 9.78{$\AA^3$} compared to 15.44{$\AA^3$} for the set of all C-type structures with 14 Co, with the largest variance in the lattice parameters being that of the largest lattice parameter, $a$. On this basis, the ability to distinguish structures with the same number of oh and td Co may be beyond the capabilities of the current model. However, the ability to distinguish structures with the same number of Co but a different td:oh ratio is still attainable, and important when later accounting for different sources of entropy \cite{foreground_paper}.

\subsubsection*{Additional tetragonal training data}

%\begin{figure}[h!]
%\includegraphics[width=0.45\textwidth]{figures/tetrag_outliers.png}
%\caption{Plot of formation energy per formula unit showing DFT calculated values for set A- and B-type structures and values predicted by machine learning (MTP) for set C-type tetragonal structures trained on only the original DFT data sets. The energy of the structures with the minimum energy (as predicted by MTP) for 11 different compositions have been re-calculated with DFT and are shown in red.}\label{MTP_outliers}
%\end{figure}

%Fig.~\ref{phase_transition}e showed all of the DFT training data for set A-, B- and C-type structures split into structures identified as cubic or tetragonal. On the 
The Co-poor side of Fig.~\ref{phase_transition}e shows coexistence between tetragonal and cubic C-type structures, with the latter corresponding to higher energies. This behaviour is consistent with the tetragonal to cubic phase transition in the pure {\MnO} compound \cite{MnO_phases}. %There are therefore both cubic and tetragonal structures for Co-poor configurations of {\alloy} in the training data.
However, there is no such phase transition in the pure {\CoO} compound.
%and Fig.~\ref{phase_transition}e does not show any phase coexistence for Co-rich C-type structures. There is therefore no data for Co-rich tetragonal structures in the training data, which could lead to extreme extrapolation by the ML model when predicting the energy of such structures. 
For tetragonal structures, beyond the Co-poor range, our method of creating structures based on estimated lattice vectors will create some tetragonal phases that are not represented in the DFT training data, which could lead to extreme extrapolation by the ML model.

Choosing a hard cut-off composition for the existence of tetragonal structures based on the available DFT training data would be a large approximation. Instead, additional tetragonal structures are calculated with DFT and added to the training set so that some of these types of structure are present in the training data. The structures were selected by training a MTP for tetragonal structures with just the available DFT training data and predicting energies for thousands of randomly generated tetragonal C-type structures. The structures at each composition 5-15 Co in the supercell that were predicted to have the lowest energy by the MTP model were then selected for the additional DFT calculations. 
%The original MTP predicted energies and the re-calculated DFT energies for the selected structures are shown in Fig.~\ref{MTP_outliers}. 
The tetragonal MTP was then re-trained with this additional data for all subsequent parts of this work.

\subsubsection*{Energy predictions when training with the estimated lattice parameters}

\begin{figure}[h!]
\includegraphics[width=0.475\textwidth]{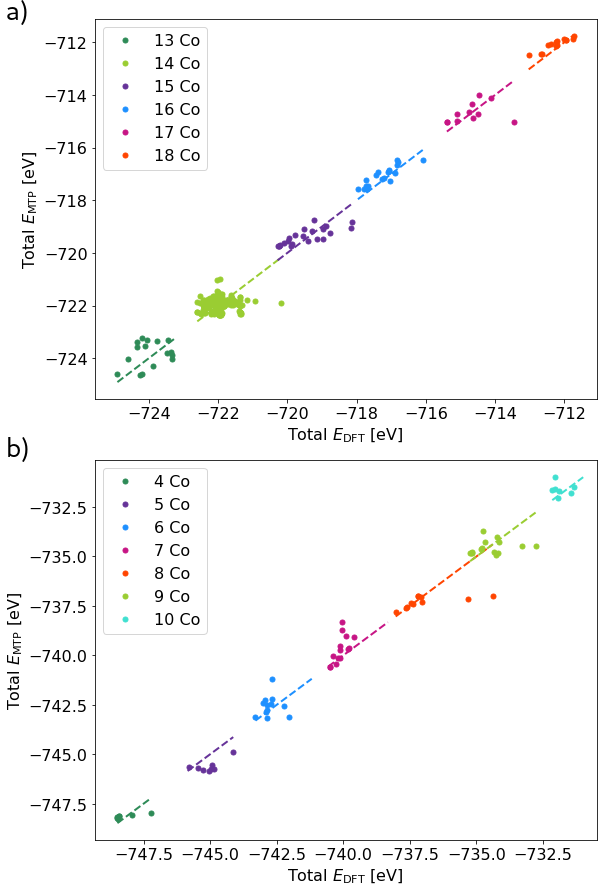}
\caption{Scatter plots for calculated DFT total energies against MTP predictions for hold-out sets of C-type structures with different total number of Co in the supercell for a) cubic structures, b) tetragonal structures.}\label{holdout_tests}
\end{figure}

To assess the level of discrimination afforded by the method,
firstly MTP was trained with all cubic data except those corresponding to $(n_{\mathrm{oh}}, n_{\mathrm{td}})=(12,2)$, which formed the hold-out set. $n_{\mathrm{td}(\mathrm{oh})}$ stand for the number of Co per supercell in td(oh) sites. This test yields an RMSE of 3.61 meV per atom, but it does not achieve good correlation between reference DFT energies and model predictions. This suggests that one would need a fully trained interatomic potential to distinguish between structures sharing the same $(n_{\mathrm{td}}, n_{\mathrm{oh}})$, which is not possible with the available amount of training data.

On the other hand, when the hold-out sets contain different $n_{\mathrm{td}}/n_{\mathrm{oh}}$ ratios, the cubic-trained MTP displays increasingly larger Pearson correlation and smaller RMSE's as the total number of Co is increased (Table~\ref{cubic_holdouts}). An exception is the 17 Co dataset, whose reduction in correlation is caused by a single outlier (Fig.~\ref{holdout_tests}a).   From approximately 16 Co in the supercell (2 per formula-unit), the minimum energy structures for all three substitution schemes are all cubic (Fig.~\ref{phase_transition}). Therefore the volume-composition relationship is much simpler and likely to be better represented by our surfaces of best fit.

For cubic structures with less than 15 Co in the supercell, correlations were typically poorer. For the 14 Co hold-out set, this is due to a large portion of the set being structures with the same oh:td ratio for which, as already discussed, the model with estimated lattice parameters does not yield well-correlated energy predictions. Furthermore, as can be seen from Fig.~\ref{phase_transition}, between the different substitution schemes there is much more variation in cell parameters at intermediate compositions where the phase transition occurs at different compositions for the different substitution schemes. Energy predictions for cubic structures at intermediate compositions are therefore likely to be less reliable than those in the `clearly cubic' regime for $>$15 Co in the supercell.

\begin{table}[]
\caption{RMSE in meV per atom and Pearson correlation coefficient between reference DFT energies and MTP predictions for cubic holdout set C-type structures with different total number of Co in the supercell.}
\label{cubic_holdouts}
\begin{tabular}{@{}lllllll@{}}
\toprule
\# Co          & 13  & 14   & 15    & 16    & 17    & 18    \\ \midrule
RMSE           & 6.06  & 2.46 & 3.57  & 1.40  & 5.40  & 1.05  \\
Pearson coeff. & 0.278 & 0.0758 & 0.787 & 0.878 & 0.313 & 0.934 \\ \bottomrule
\end{tabular}
\end{table}

\begin{table}[]
\caption{RMSE in meV per atom and Pearson correlation coefficient between reference DFT energies and MTP predictions for tetragonal hold-out set C-type structures with different total number of Co in the supercell.}
\label{tetrag_holdouts}
\begin{tabular}{@{}lllllllll@{}}
\toprule
\# Co          & 4     & 5    & 6     & 7     & 8     & 9     & 10     \\ \midrule
RMSE           & 2.38  & 7.51 & 7.52 & 8.29  & 16.9  & 8.58  & 4.63   \\
Pearson coeff. & 0.889 & 0.726 & 0.0160 & 0.661 & 0.597 & 0.249 & -0.133 \\ \bottomrule
\end{tabular}
\end{table}

Similar hold-out tests were also performed for tetragonal C-type structures and a MTP trained only on tetragonal data (Fig.~\ref{holdout_tests}b and Table~\ref{tetrag_holdouts}). In this case, better correlation was typically achieved for Co-poor structures, i.e. in the `more tetragonal' regime. However tetragonal structures have considerably more variation in lattice parameters than cubic ones, even for as few as 4 Co in the supercell in set B (Fig.~\ref{phase_transition}b). It is therefore likely that the lattice parameter estimates are less accurate for tetragonal structures, especially those compositionally closer to being B-type. This is reflected by the relatively large RMSE's presented in Table~\ref{tetrag_holdouts}. The correlation substantially decreases upon increasing number of Co in the supercell. 

Intermediate compositions have a complex composition-dependent phase coexistence and substantial cell volume variation. There, accurate energy predictions likely can only be obtained using a trained full potential capable of performing structural relaxation. However, for very Co-poor structures and those with $>$15 Co in the supercell, energy predictions with estimated lattice parameters are expected to be more reliable.
In all hold-out set tests performed in this section, the settings for MTP training that achieved the best RMSE were the level-eight (i.e. with $\mathrm{lev}_{\mathrm{max}} = 8$, see Section~\ref{MTP_overview}) with weights of 1.0 for energy and 0.0 for stress and forces (due to lack of force information in initial training data). The cutoff radius was $R_{\rm cut} = 5$\AA. These settings were therefore used in all subsequent parts of this work for sampling the configuration space of set C-type structures.

\subsubsection{Filling in the configuration space}

\subsubsection*{Random sampling of set C}

\begin{figure*}[ht]
\includegraphics[width=1.0\textwidth]{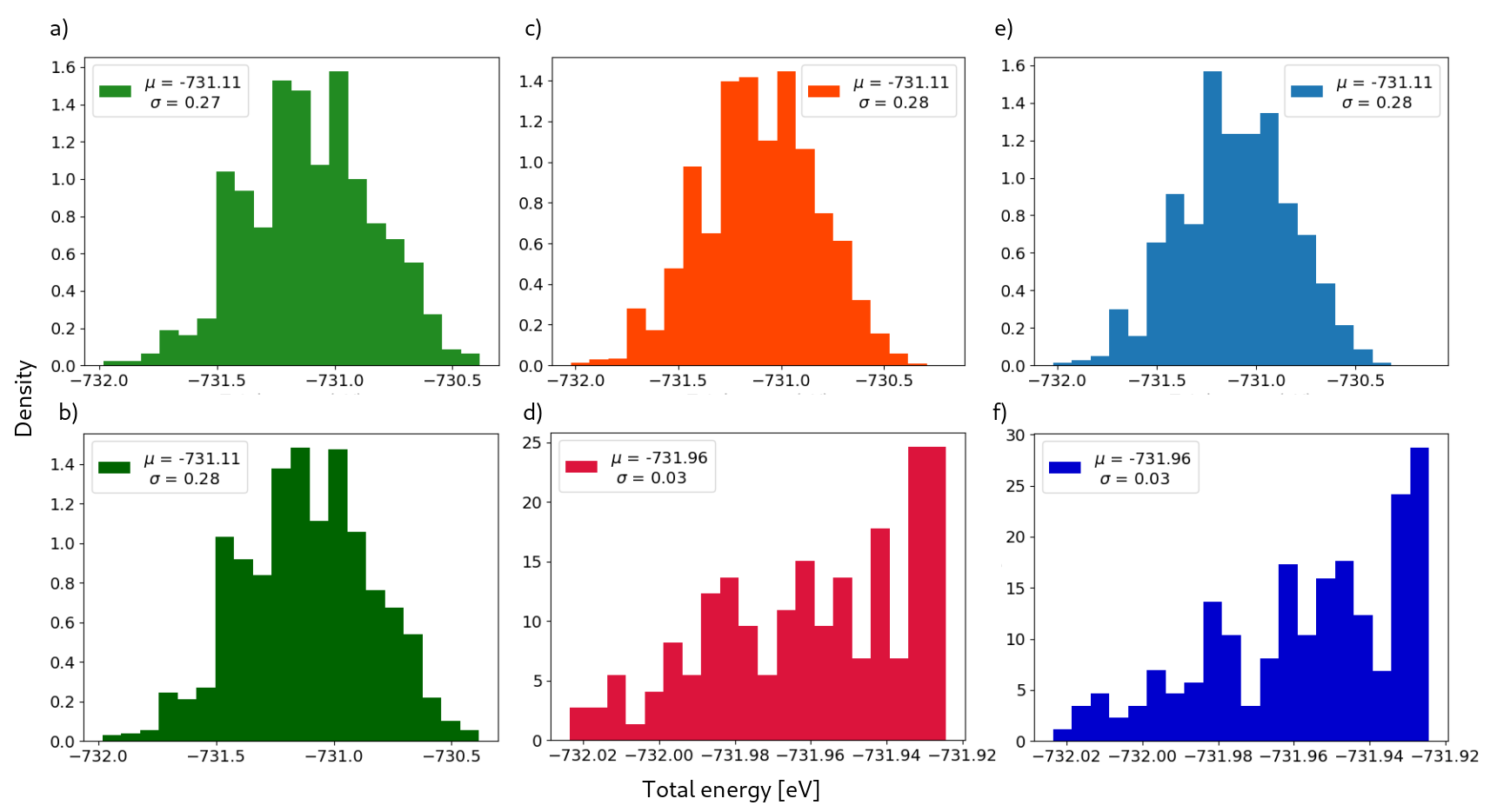}
\caption{MTP total energy predictions for random samples of set C type structures of {\alloy} with 10 Co in the supercell. a) 1k attempts batch. b) Batch from (a) augmented with a second 1k batch. c) 100k attempts batch. d) Lowest 0.1 eV of distribution in (c). e) 2 million attempts batch. f) Lowest 0.1 eV of distribution in (e). The mean ($\mu$) and standard deviation ($\sigma$) of each distribution is given in the legend.}\label{MC_sampling}
\end{figure*}

To sample the millions of possible configurations of the 56-atom supercell for C-type structures, we generate random structures and only retain unique configurations. Fig.~\ref{MC_sampling} compares the distribution of total energies for configurations with 10 Co in the supercell predicted by MTP for random batches of 1,000, 100,000 and 1,949,176 unique structures (the limit of possible configurations for this particular composition is 1,961,256).
The lowest energy side of the distributions deserves particular attention, as these configurations contribute the most to the total free energy (Fig.~\ref{MC_sampling}d and f). The mean and standard deviation of the distribution for the sample of 100,000 structures (Fig.~\ref{MC_sampling}c and d) has converged to that of the (almost) complete sample (Fig.~\ref{MC_sampling}e and f). Therefore, for all compositions of set C, the generation of 100,000 unique structures is attempted (noting that at some compositions there are not as many as 100,000 unique structures in the total configuration space). Total energies are then predicted for these structures with MTP and EMF models to sample the full composition space of set C.

\subsubsection*{Recovering the symmetry degeneracy of the training data}

During the generation of the training data for set A- and B-type structures in Ref.~\citenum{Javi_paper}, the software package CASM \cite{CASM} was used to select only symmetrically unique structures in order to reduce the total number of necessary DFT calculations. %This is not the case for the set C structures generated randomly in this study. 
However, this introduces a bias into the data.
A similar approach is adopted by the SOD software package \cite{SOD} to reduce the number of calculations to perform when modelling disordered solids, but in this case the degeneracy is retained to allow the computation of `entropy-reduced' energy.
For the set A- and B-type training data, it was necessary to first recover the symmetry degeneracy to remove the bias when sampling the configuration space and we outline our procedure for this in the SI (Section 5).
%Refer to method in SOD \cite{SOD} of retaining degeneracy when scanning a list of pre-defined structures to compute a `entropy-reduced' energy.

%Old way: Cite use of spglib (cite). Make flow chart to outline algorithm used here and list all key assumptions made? E.g. threshold used by spglib, using initial cubic structure (most closely mimics initial procedure performed by CASM to generate training data), generating a `fractional degeneracy' to allow us to compare structures and weight appropriately when scaling by combination space. See symmetry algorithm notebook for further details on algorithm, assumptions and possible limitations/ consequences of assumptions made.

%Various methods have been proposed in the literature to obtain an unbiased configurational sampling. Here the SOD method \cite{SOD} is adopted, but with their original method to determine structure equivalence replaced by an alternative one based on atomic environment descriptors. SOD's approach to tell if two structures are the same involves applying all the symmetry operations of the parent structure onto one of them, and checking if it is turned into the other one. Instead, 

\subsubsection*{Scaling data by combinatorial space}
To scale by the total number of possible combinatorial substitutions, the data is grouped by total number of Co atoms in the supercell and then by the total number of Co on oh sites. A scaling factor is then determined by finding the factor necessary to scale up the total group size to be equal to the total combination space for that particular number of Co and Co-on-oh-sites,
\begin{equation}\label{comb_space}
    c_{\mathrm{tot}} = \frac{n^{\mathrm{td}}_{\mathrm{tot}}!}{n^{\mathrm{td}}_{\mathrm{Co}}!( n^{\mathrm{td}}_{\mathrm{tot}} - n^{\mathrm{td}}_{\mathrm{Co}})!} \times \frac{n^{\mathrm{oh}}_{\mathrm{tot}}!}{n^{\mathrm{oh}}_{\mathrm{Co}}!( n^{\mathrm{oh}}_{\mathrm{tot}} - n^{\mathrm{oh}}_{\mathrm{Co}})!},
\end{equation}
where $n^{\mathrm{td}}_{\mathrm{tot}}$ is the total number of td sites in the 56 atom supercell, $n^{\mathrm{td}}_{\mathrm{Co}}$ is the number of the td sites occupied by Co. Similarly, $n^{\mathrm{oh}}_{\mathrm{tot}}$ is the total number of oh sites and $n^{\mathrm{oh}}_{\mathrm{Co}}$ is the number of these sites occupied by Co.

For the randomly generated set C structures, this simply results in multiplying each structure by the scaling factor. However for set A and B, which were set up to be symmetrically distinct structures, instead the symmetry degeneracies of all structures with the same Co count and Co-on-oh-sites count are summed to determine the total group size for each Co count and Co-on-oh-sites count. The symmetry degeneracies are then scaled so that the total symmetry degeneracy for each group equals the total combination space for the particular number of Co and Co-on-oh-sites. This preserves the weight of each structure in set A and B, where some structures would have had more equivalent structures if they had been generated randomly. Note that not all configurations generated by CASM were successfully relaxed, but by scaling the structures that did relax by configuration space, essentially an average is being taken over the structures that did relax as an approximation for missing structures from the training set.

%\section{Results \& discussion}

\subsection{Prediction of free energies}\label{free_energy_pred}

\begin{figure*}[ht]
\includegraphics[width=1.0\textwidth]{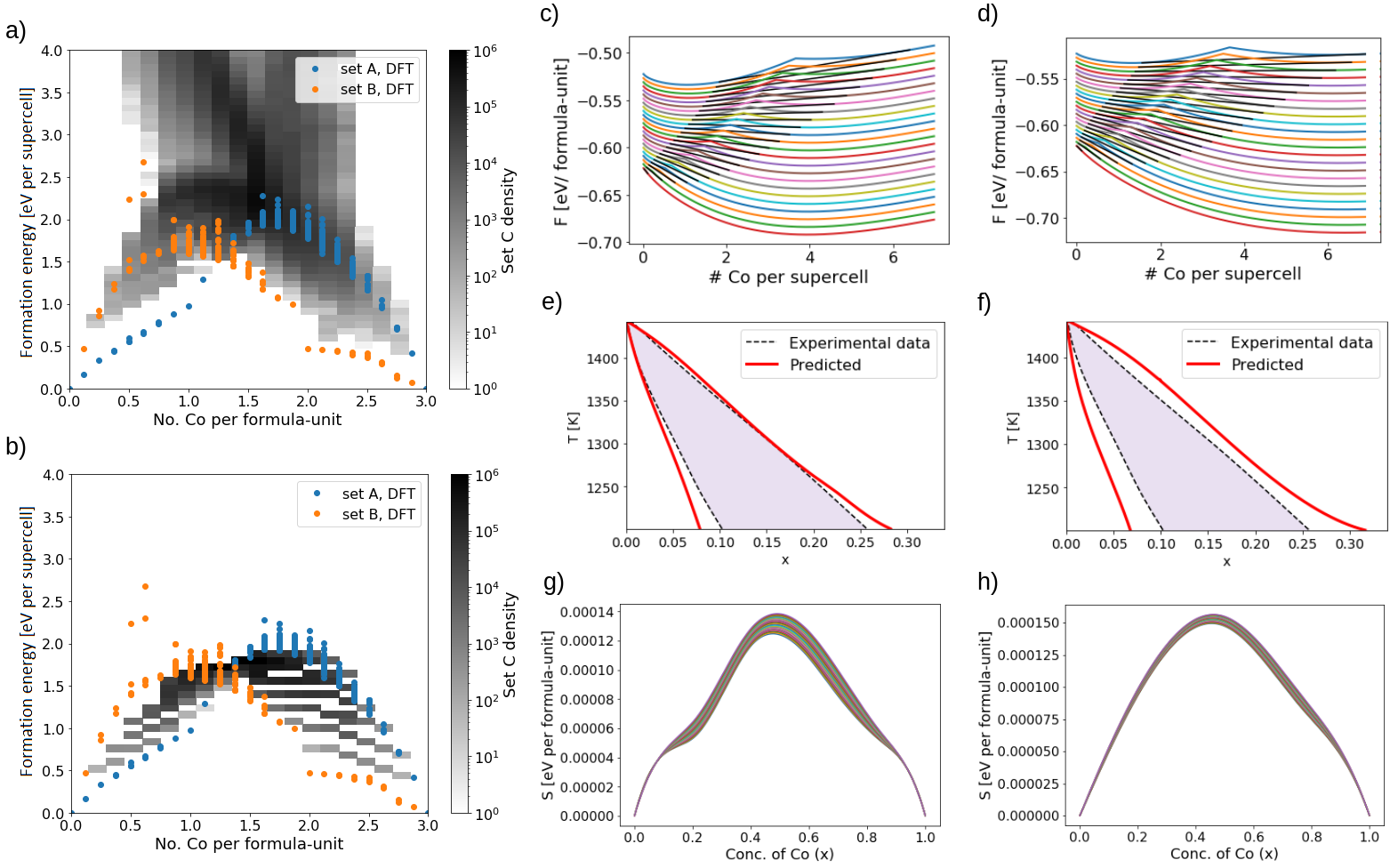}
\caption{Formation energy as a function of Co composition in supercells of {\alloy} with set C predicted energies shown in black from a) the MTP model and b) the EMF model. Free energy curves calculated across a temperature range of 1200-1440 K including: magnetic, vibrational and configurational entropy as obtained from c) the MTP model sampling of set C and d) that of the EMF model. Solubility gaps obtained from the tangents of the free energy curves (black lines in (c) and (d)) are shown in (e) and (f) respectively, where experimental data is taken from Ref.~\citenum{CoMnO_expt}.  Configurational entropy extracted from each model is shown in g) for MTP and h) for EMF.}\label{main_results}
\end{figure*}

%External conditions such as temperature and overall composition determine material properties such as separation into phases, atomic level ordering, or the distribution of species amongst inequivalent sites. These structural properties can be probed with experimental techniques, and they allow us to validate our theoretical models. In {\alloy}, we have the spindle-shaped solubility gap (SG) between the cubic and tetragonal phases on the Co-poor side \cite{CoMnO_expt}. On the Co-rich side there is no such feature, but on the other hand, there are techniques to look at how the Co distributes amongst the site-types (td or oh). Below we show how to calculate the free energy and discuss the thermodynamic properties that arise from it as predicted with the MTP and EMF methods.

The thermodynamic stability of alloys, or other mixed phases such as {\alloy}, depends on minimising the Gibbs free energy,
\begin{equation}\label{Gibbs}
    G(N, P, T; \vec{x}) = H - TS ,
\end{equation}
where $N$ is the number of atoms, $P$ is the pressure, $T$ is the temperature, $H$ is the enthalpy, $S$ is the entropy and $\vec{x}$ is a vector representing the full set of molar fractions of the alloying species \cite{Widom_rev}. 
%The thermodynamic stability of a particular composition also depends on whether it has a lower free energy than that of any competing phases. For this reason, it can become unfavourable to add too many additional alloying species due to the increased number of possible competing phases \cite{Widom_rev}. The Gibbs free energy can also be expressed as,
%\begin{equation}\label{Gibbs2}
%    G(T, \vec{x}) = \mathrm{min}_V (F + PV)
%\end{equation}
%where $F$ is the Helmholtz free energy and $V$ is the volume. The synthesis of {\alloy} is often carried out at atmospheric pressure \cite{Javi_paper}. 
For solids at atmospheric pressure, $G$ can be approximated to the Helmholtz free energy, $F$ \cite{vib_entropy, Widom_stability}.
%the product $PV$ is generally negligible \cite{vib_entropy}. Therefore, the Gibbs free energy, $G$, of each phase can be approximated to the Helmholtz free energy, $F$, at its minimising volume \cite{Widom_stability}.
As shown in Eq.~\ref{Gibbs}, the free energy is reduced by $S$. 
%For this reason, methods exist to compute the `reduced energy' for disordered materials based on a calculated entropic term \cite{SOD}. 
At low temperatures, the product of $TS$ is small, so the magnitude of the free energy is dominated by $H$. However, at higher temperatures, there is a stronger reduction of the free energy by the entropic term. 

There are many different possible contributions to the entropic term, such as: the different chemical substitutions within the alloy ($S_{\mathrm{chem}}$), vibrational entropy ($S_{\mathrm{vib}}$), degrees of freedom due to electronic and magnetic excitations ($S_{\mathrm{elec}}$ and $S_{\mathrm{mag}}$ respectively) \cite{Widom_rev}, giving
\begin{equation}\label{entropy}
    S = S_{\mathrm{chem}} + S_{\mathrm{vib}} + S_{\mathrm{elec}} + S_{\mathrm{mag}} + ...
\end{equation}
From a statistical-mechanics perspective, the reduction in the free energy by an increased number of states of the system (from various entropic contributions) can be understood as the phase enclosing more states in its phase space being more likely to be visited as the system undergoes microscopic transitions and hence has an increased stability relative to other phases \cite{vib_entropy}. 
The stabilisation of multicomponent alloys due to the entropy of mixing from chemical substitution is a fundamental concept of high entropy alloys (HEAs) and there are a number of reviews on this particular subject such as Ref.~\citenum{Widom_rev}, \citenum{HEA_rev} and \citenum{HEA_rev2}.
{\alloy} does not technically meet the specifications to be considered a HEA, such as containing five or more elements in nearly equal atomic ratios \cite{HEA_rev2}. However, this does not eliminate the possibility of stabilisation of this system from configurational or other entropy sources.

All available training data was used to train a MTP and EMF model, where for the MTP model the training data was split into tetragonal and cubic structures. While our EMF model takes only number of oh and td Co in the supercell as inputs for describing the system, it was necessary to estimate lattice parameters for all of the randomly generated structures used to sample the configuration space of set C with the MTP model, as described earlier. 
The EMF model involved 10 fitting parameters compared to 156 for the MTP model when training with the 8g basis set and 3 different atomic species. These trained models were used to predict the energies of the full composition space of set C using the randomly generated structures and scaling by combination space. 
As the EMF model is unable to distinguish structures with the same number of oh and td Co in the supercell, but with different cell shapes and energies, only the minimum energy structure in the training set for each number of oh and td Co is used to train the model. When predicting the free energies with the EMF model, the scaling factor for set C structures is divided by two as with the EMF model there is not a cubic and tetragonal version of the same structure. With this model, there is only the minimum energy phase at each composition.

Total energies for C-type configurations of {\alloy} predicted by the MTP and EMF models were used to produce plots of formation energy vs. composition via Eq.~\ref{form_E}, which are shown in Fig.~\ref{main_results}a and b for MTP and EMF models respectively. It can be seen that the EMF model produces much more discretised energy predictions due to its inability to distinguish structures with the same number of oh and td Co. These formation energies were used to calculate $F$ as a function of temperature and Co concentration, $x$, including also magnetic entropy, vibrational entropy and a correction to the configurational entropy at the composition edges (due to finite-size limitations of the supercells). These methods are outlined in Ref.~\citenum{foreground_paper}, where the importance of each of these contributions was demonstrated. Calculated $F$ curves from MTP and EMF are shown in Fig.~\ref{main_results}c and d respectively.

The calculated $F$ from each model is then used to compute the SG (or phase coexistence region) between tetragonal (H) and cubic (S) phases of {\alloy}. This method is again described in Ref.~\citenum{foreground_paper}. Calculated SG's from MTP and EMF are shown in Fig.~\ref{main_results}e and f respectively and compared to experimental data from Ref.~\citenum{CoMnO_expt}.
Both methods gave similar SG's in this low-Co concentration range, but the EMF model predictions diverge more from the experimental SG, especially at very Co-poor compositions and higher temperatures. 
%This may be due to the discretisation of energy predictions with the EMF model and hence more error at lower temperatures when states are not energetically accessible, which perhaps should be at the given temperature. 

Fig.~\ref{main_results}g and h show the extracted configurational entropy from MTP and EMF models respectively. This is extracted from the predicted configurational free energy, $F_{\mathrm{config}}$, via,
\begin{equation} \label{entropy_Landau}
    S_{\mathrm{config}} = - \Bigg(  \frac{ \partial F_{\mathrm{config}} }{\partial T} \Bigg)_V .
\end{equation}
The most noticeable difference between the configurational entropy predictions by the two models is the Co-poor side where the EMF model gives larger configurational entropies than with the MTP model at the same Co concentrations.

Based on the better agreement between the MTP prediction and experimental data for the SG in the Co-poor region of the phase diagram, and from comparing the set C energy predictions to the DFT data for C-type structures (shown in Fig.~\ref{phase_transition}e), the MTP predictions appear to reproduce the known features of the system more accurately. One may also ask how classical force fields would perform versus ML predictions. However, it is hard to make a comparison between classical force fields and machine-learned potentials on an equal footing because of the different methodologies and philosophies of their development: the classical potentials are typically developed through many loops of trial-and-error, while the machine-learned potentials are trained automatically on an ab initio database generated ad hoc for the present problem. Some explicit comparisons can be found in refs.~\cite{boes_neural_2016,grabowski_ab_2019}.

With the experimental data available for the H+S SG, we are only able to investigate the accuracy in the calculation of the phase diagram, and the different ML methods used to sample set C, for the Co-poor side of the phase diagram, up to approximately $x$ = 0.3 in {\alloy}. Further experimental data for Co-rich phases would provide valuable information for assessing the accuracy of the ML predictions, similar to the cation site occupancy measurements performed in Ref.~\citenum{CoMn2O4_XRD} for a Mn-rich composition. It was demonstrated in Ref.~\citenum{foreground_paper} that composition-dependent magnetic entropy and, in particular, the vibrational entropy from a higher energy cubic phase are vital to accurately reproduce the SG in the Co-poor region. However, on the Co-rich side, where there is no such higher energy phase, vibrational entropy may be less dominant in determining $F$ and hence other factors such as the accuracy in the sampling of C-type configurations (which appear to be closer in energy to the ground state on the Co-rich side in Fig.~\ref{main_results}a) may play a stronger role in the calculation of $F$. In the SI (section 6) we compare the dominant set structure type as a function of temperature and Co composition as predicted by the EMF and MTP models.

\section{Summary \& further work}

We have explored the use of three different ML methods to sample the full configuration space of {\alloy}: ANNs and MTPs, originally developed to implement ML potentials; and a much simpler EMF method, requiring as inputs only the number of substituting species occupying particular types of crystallographic site (i.e. octahedral or tetrahedral in the case of {\alloy}).

Our ANN model was capable of achieving some of the lowest validation RMSE's in many test cases. However, in an imbalanced data set, as in the case of {\alloy} with fewer possible configurations at the composition extremes without any data pre-treatment, the ANN performed more poorly than the other methods on minority type data. For the MTP model, due to the lack of force information in our initial data set, it was necessary to obtain estimates for the relaxed lattice parameters of the set C structures based on the DFT training data. In the Co-poor composition range of the phase diagram we have compared available experimental data \cite{CoMnO_expt} to model predictions. In this composition range both EMF and MTP models provide reasonably accurate predictions, with the MTP predictions being superior to those of the EMF model at the lowest Co concentrations and highest temperatures.

Studying the Co-rich side of the phase diagram of {\alloy} and the spinel-rocksalt phase coexistence region \cite{CoMnO_expt} could provide valuable insights for energy storage via redox reactions \cite{Javi_paper, mixed_oxides, Andr2018}. To study this part of the phase diagram, further experimental data to check the accuracy of ML energy predictions would be very valuable. For example, the predicted relative occupation of td and oh sites by Co atoms at different temperatures and concentrations becomes non-trivial beyond 8 atoms per supercell: as we show in the supplementary materials, MTP and EMF models predict slightly different occupations, which could be experimentally verified to assess the validity of the models~\cite{supplemental-information}. Furthermore, it would be necessary to have additional \textit{ab initio} data for rocksalt structures. It may also be important to consider the effect of a possible partial low-to-high spin-state transition of oh Co in {\CoO} \cite{Co3O4_spin1, Co3O4_spin2, Co3O4_spin3, Co3O4_spin4}, which we have neglected in this work as it focused on the Co-poor side of the phase diagram, but could impact the magnetic entropy for Co-rich structures.

\section*{Data access statement}
Software developed in this work for calculating the solubility gap of {\alloy} with the EMF model and with data from the MTP model in this work is available from \href{https://doi.org/10.5281/zenodo.4133866}{https://doi.org/10.5281/zenodo.4133866} under a BSD 3-Clause license.
Data from \textit{ab initio} calculations used to train machine learning models in this study is available from \href{https://dx.doi.org/10.17172/NOMAD/2020.10.14-1}{https://dx.doi.org/10.17172/NOMAD/2020.10.14-1}. The trained MTP potential files are also provided in the SI.\\

\section*{Acknowledgements}
We thank Herv\'{e} Manzanarez for discussions during the development of our mean field model.
Work at CEA-Grenoble was supported by Institut Carnot, through projects MAPPE and PREDICT. This work was performed using HPC resources from GENCI-TGCC Grant A0060910765. 
J. C. acknowledges the computer resources at SGI/IZO-SGIker UPV/EHU, the i2BASQUE academic network, and MareNostrum and the technical support provided by the Barcelona Supercomputer Center (grant No. QS-2020-2-0002). Work at CIC energiGUNE was supported by the Ministerio de Ciencia e Innovación of Spain through the project ION-SELF (No. PID2019-106519RB-I00). 
A.S. was supported by the Russian Foundation for Basic Research under Grant No. 20-53-12012

\bibliography{alloyRefs}% Produces the bibliography via BibTeX.

\end{document}